\documentclass[final,5p,times]{elsarticle}  

\usepackage{microtype}        
\usepackage[utf8]{inputenc}   
\usepackage{amsmath,amssymb,amsfonts}
\usepackage{xcolor}
\usepackage{transparent}  

\usepackage{graphicx}         
\usepackage{subcaption}        
\usepackage[colorlinks=true,allcolors=blue]{hyperref}
\usepackage[capitalise,nameinlink]{cleveref}
\usepackage{orcidlink} 
\usepackage{stfloats} 

\usepackage[switch]{lineno}

\journal{NIM A preprint}

\begin{document}

\begin{frontmatter}


\title{Mu2e Straw Tube Tracker Gas Flow Quality Control}


\author[label1]{V.~Bharatwaj}
\author[label3]{S.~N.~Israel}
\author[label2]{M.~Jangra\orcidlink{0000-0002-7766-1718}}
\author[label6,label1]{T.~M.~Nguyen\orcidlink{0000-0002-9978-7813}}
\author[label1]{J.~M.~Peck\orcidlink{0009-0007-7043-9819}}
\author[label4]{M.~Stortini}
\author[label3]{N.~H.~Tran\orcidlink{0000-0002-5242-6690}}

\author[label2]{D.~Ambrose\orcidlink{0000-0003-3779-2702}}
\author[label1]{A.~Edmonds\orcidlink{0000-0002-8522-1368}}
\author[label5,label2]{H.~Hass}
\author[label2]{E.~R.~Martin}
\author[label2]{A.~Mukherjee\orcidlink{0000-0001-5151-4333}}
\author[label5]{K.~Northrup}
\author[label1]{J.~L.~Popp\orcidlink{0000-0002-5415-0800}}
\author[label2]{V.~L.~Rusu\orcidlink{0009-0005-4224-0605}}
\author[label2]{R.~S.~Tschirhart}
\author[label2]{R.~L.~Wagner}

\author{\\ On behalf of the Mu2e Collaboration}

\affiliation[label1]{organization={York College, City University of New York},
            city={Jamaica},
            postcode={11451}, 
            state={NY},
            country={USA}}
\affiliation[label3]{organization={Boston University},
            city={Boston},
            postcode={02215}, 
            state={MA},
            country={USA}}    
\affiliation[label2]{organization={Fermi National Accelerator Laboratory},
            city={Batavia},
            postcode={60510-5011}, 
            state={IL},
            country={USA}}
\affiliation[label6]{organization={FPT University, Danang Campus},
            city={Danang},
            state={Vietnam}
            }        
\affiliation[label4]{organization={Yale University},
            city={New Haven},
            postcode={06520}, 
            state={CT},
            country={USA}}  
\affiliation[label5]{organization={University of Minnesota},
            city={Minneapolis},
            postcode={55455}, 
            state={MN},
            country={USA}}

\begin{abstract}
We present a tracker gas flow quality control method developed for the Mu2e straw tube tracker. Using time-dependent current measurements, we quantify the onset time of ionization gain induced by an \(\mbox{}^{55}\mathrm{Fe}\) source during gas exchange, which is correlated to the gas conductance in the straw. This allows for the identification of channels with inadequate flow. This approach is broadly applicable to other gaseous detectors that require high-channel-count screening.\\ 
\end{abstract}



\begin{keyword}

Mu2e \sep straw tube \sep gas \sep flow \sep gain rise-time
\end{keyword}

\end{frontmatter}



\section{Introduction}

The Mu2e Experiment at Fermi National Accelerator Laboratory (Fermilab)~\cite{mu2e} will test charged lepton flavor violation (CLFV) by searching for coherent muon$-$to$-$electron ($\mu\text{-}e$) conversion, a process that has not yet been observed.
The experiment searches for $\mu$-e conversion from stopped muons in an aluminum target ($\mu^{-}\mathrm{Al}\rightarrow e^{-}\,\mathrm{Al}$), using a straw tube tracker that provides coordinate measurements of charged particles utilized in the reconstruction process.
The conversion signal is expected to appear as a monoenergetic peak at \(104.9\,\text{MeV}\), at the endpoint of the background decay-in-orbit electron spectrum.
Therefore, a critical requirement for Mu2e is precision momentum reconstruction, which the straw tube tracker provides.

\begin{figure}[!b]
  \centering
  \includegraphics[width=0.3\textwidth]{ 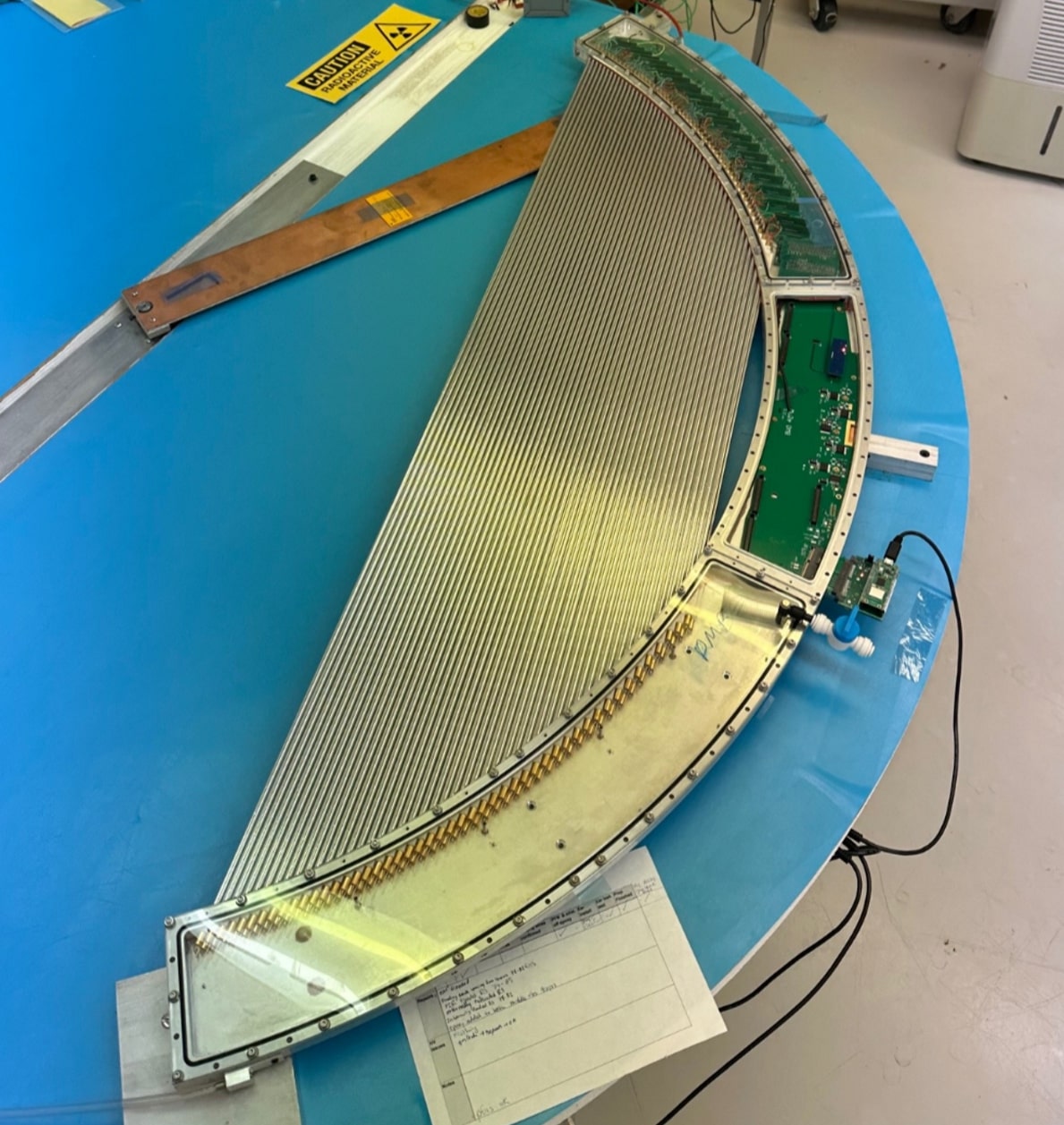}
  \caption{\label{fig:Fe55real}Photo showing an overview of a panel test setup. The temporary cover with the auxiliary valve is on the inlet side, the Digital Mother Board for data acquisition, and current amplifiers on the outlet side.}
\end{figure}

It is essential to evaluate the performance of the mu2e straw tracker. 
The method employed measures the gain response time by sweeping an \(\mbox{}^{55}\mathrm{Fe}\) source during gas exchange to identify blockages.
Other metrics that quantify gas conductance, such as current plateau values, can be measured to provide a complementary analysis.
The requirements for this test are to quickly identify potential gas-flow issues in over 20,000 straws during the assembly process.
Gas flow through individual straws is crucial to the tracker's performance, where insufficient flow impacts the gain and compromises detection.
Individual flow restrictions are difficult to detect and repair due to the tightly packed straw mounting geometry.

Previous quality-control tests of straw tube trackers commonly use the current response from a radioactive source and are designed to investigate a range of performance issues.
For example, in the muon g-2 experiment at Fermilab, the straw tube tracker modules were also tested using a radioactive source on a movable stage to assess the detector resolution using the source position relative to the mean drift time~\cite{gm2}.
Similarly, the ATLAS experiment at CERN used a stationary radioactive source while exchanging various gases to verify straw tube flow and also measure the effects of aging~\cite{atlas}.
Finally, the LHCb Outer Tracker performed scans of completed modules with a radioactive source to verify the channel response after production and installation~\cite{lhcb}.

This article describes a quality control method for identifying straws with compromised gas flow.
Section~2 describes the Mu2e tracker panel test setup; Section~3 details the gas flow QC procedure and data analysis; Section~4 provides the conclusions.

\section{The Mu2e Tracker Panel Test Setup}

\begin{figure*}[t]
  \centering
  \includegraphics[width=0.6\textwidth]{ 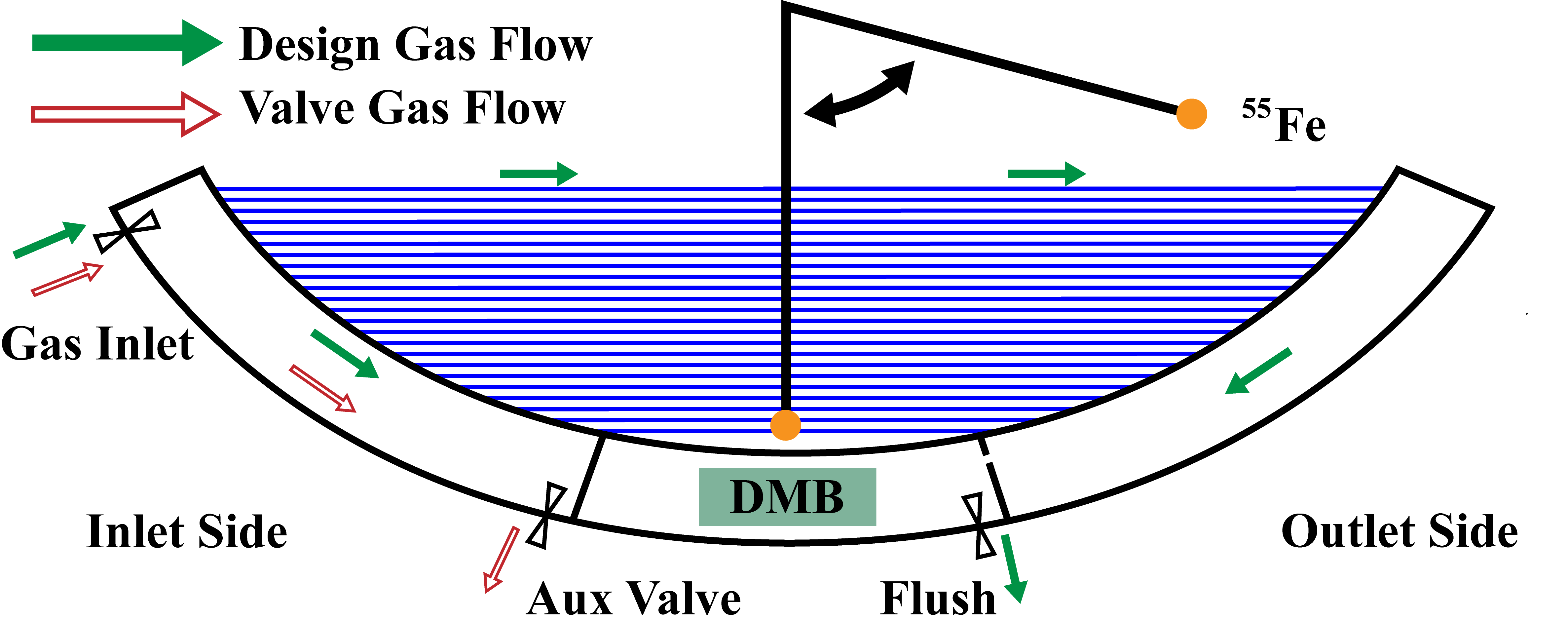}
  \caption{\label{fig:Fe55-openvalve} Illustration of the panel test setup with gas flow pattern. The design pattern, with the auxiliary valve closed, is indicated by solid arrows. The valve pattern, with the auxiliary valve open, is shown using outlined arrows.}
\end{figure*}


\begin{figure*}[b]
\centering
  \begin{subfigure}{0.3\textwidth}
    \centering
    \includegraphics[width=0.9\linewidth]{ 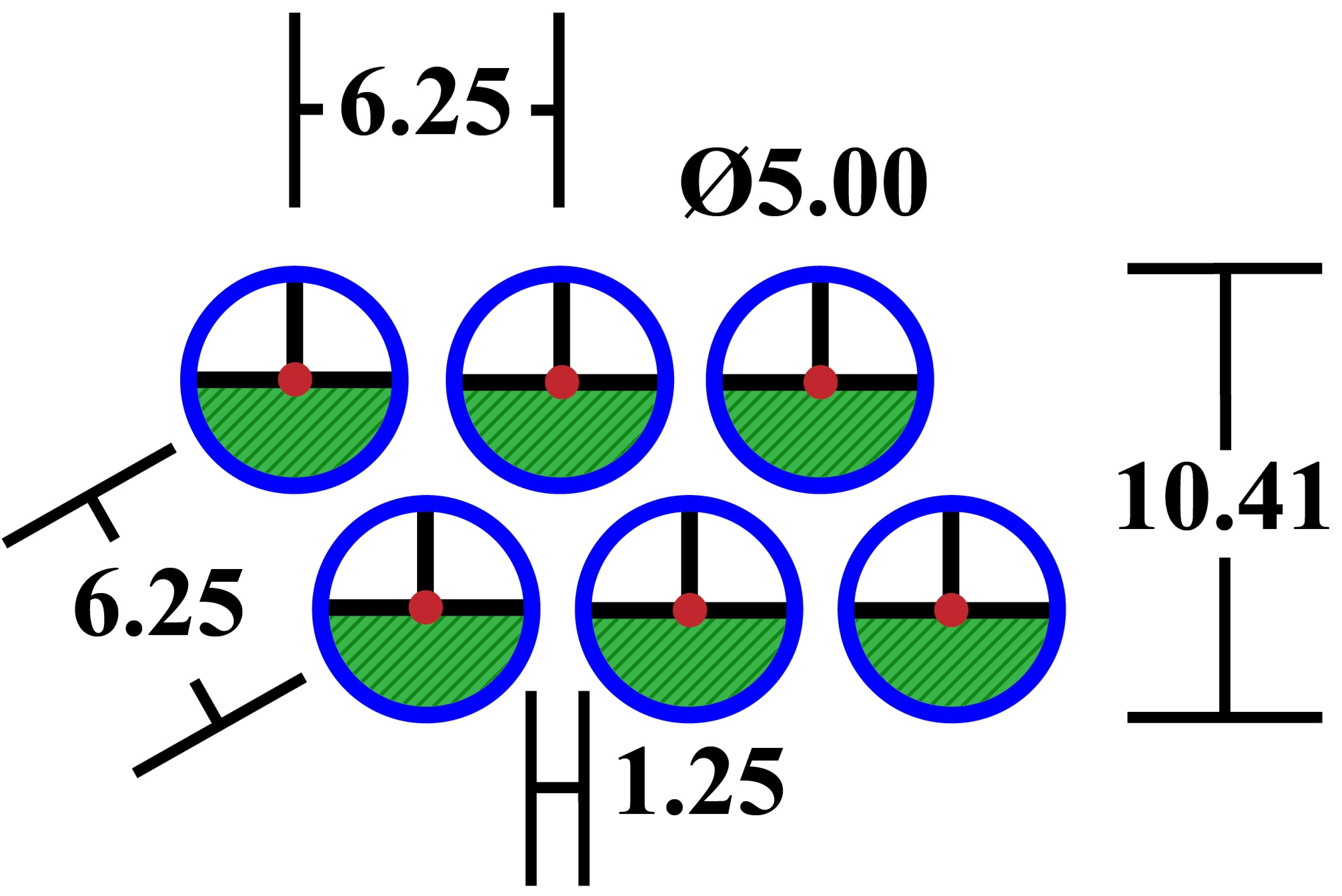}
    \caption{\label{fig:strawlayout}} 
  \end{subfigure}
  \begin{subfigure}{0.65\textwidth}
    \centering
    \includegraphics[width=0.9\linewidth]{ 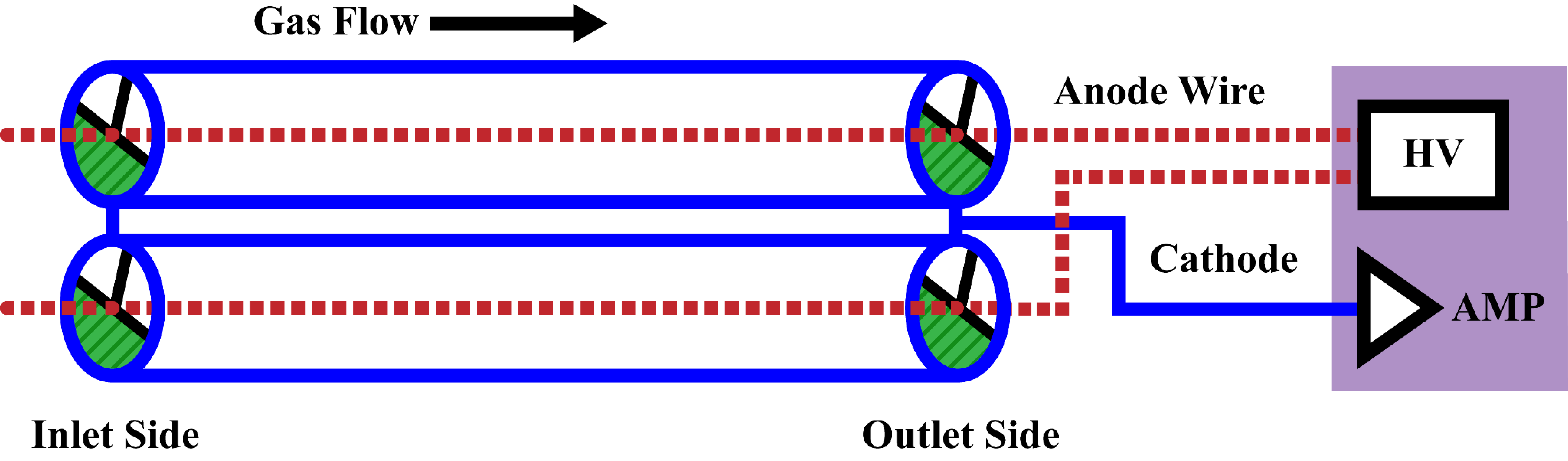}
    \caption{\label{fig:strawtubes}} 
  \end{subfigure}
   \caption{\label{fig:strawsall} Illustrations of straw tubes. (a) Section view of the straw tube layout with measurements shown in millimeters. (b) A Straw doublet connected to a current amplifier. The cathode consists of straw tubes and their electrical connections, while the dashed lines at the center of the straw tubes represent the anode wires. The flow channels are represented by the two empty regions at the ends of the straws. The structures at the ends of the straw support both the straw and the wire. The actual terminations do not appear here. The shaded box contains the high-voltage supply and the current amplifier.} 
\end{figure*}

The straw tube tracker consists of 216 identical tracker panels \cite{mu2e}.
A tracker panel, shown in Figure~\ref{fig:Fe55real}, is divided into three gas volumes which house the on-board electronics and guide gas flow.
Referring to Figure~\ref{fig:Fe55-openvalve}, these are the inlet side volume on the left side, the outlet side volume on the right side, and the Digital~Mother~Board (DMB) volume at the center.
The solid arrows indicate the design gas flow path: gas enters through the gas inlet at the edge of the inlet side, flows through the straws to the outlet side, and finally exits the panel through the DMB volume.

Every panel consists of 96 aluminized Mylar straw tubes, arranged in two layers of 48 tubes each. 
The layers are offset so that the center of one layer aligns with the gaps of the other, as illustrated in Figure~\ref{fig:strawlayout}, to improve coverage.
The straws have a \(5\,\text{mm}\) diameter with \(15\,\mu\text{m}\) walls to reduce energy loss and multiple scattering. 
Straw lengths range from \(430\,\text{mm}\) to \(1200\,\text{mm}\) with an average length of \(910\,\text{mm}\). 
Each straw is attached to the panel with epoxied terminations that provide structural support, seal the gas volume, and allow for electrical connections.

For readout, adjacent straws from each layer are paired electrically into ``doublets'', meaning the two straws share a single output channel.
The connections for each doublet are illustrated in Figure~\ref{fig:strawtubes}.
The test current amplifier boards supply high voltage, typically around 1450~V, and measure current for each straw doublet.
Signals are then sent to the DMB for data acquisition.

During panel production, the $\sim\!2$~mm gas aperture holes in the straw terminations can be partially or entirely blocked by the epoxy compound used to secure them. 
Flow restrictions may occur at either end of the straw, potentially leading to aging effects (straw damage) in the presence of radiation, as explained in~\cite{wireaging}.
A completely blocked straw has zero gain, reducing the tracker's efficiency.
Straws with insufficient flow need to be identified and, where possible, repaired.


\section{Straw Tube Gas Flow Quality Control}

The straw tube gas flow quality control testing method for Mu2e tracker panels was developed at Fermilab.
The test setup shown in Figure~\ref{fig:Fe55real} features a panel secured on both sides with a \(0.181\,\mu\text{Ci}\) \(\mbox{}^{55}\mathrm{Fe}\) source mounted on a custom designed motorized arm driven by a stepper motor.
During the test, the motor is powered to slowly sweep the source underneath the panel at a rate of approximately one sweep per minute.
When the operating voltage is applied to each wire, the \(5.90\,\text{keV}\) and \(6.49\,\text{keV}\) X-rays produced by the \(\mbox{}^{55}\mathrm{Fe}\) can be detected~\cite{ironsource}.
The cathode current for each doublet is recorded at a sampling rate of \(10\,\text{Hz}\).
During the gas exchange process, the source passes by each straw multiple times, probing the gas gain at regular intervals.

A summary of the gas procedure is as follows:
To begin the test, the operational gas mixture ${\rm Ar-CO}_2$~(80:20) is introduced at a rate of \(14.16\,\text{NL/h}\) and an average gas pressure of one~atmosphere.
With the auxiliary valve closed, the operating gas mixture flows through the straws from the inlet side to the outlet side, as indicated by the solid arrows in Figure~\ref{fig:Fe55-openvalve}.
Typically, one hour is allowed for the ${\rm Ar-CO}_2$ to fill the panel.

Once functional doublet readout is established, the panel is flushed with ${\rm N}_2$, an inexpensive and readily available gas with very low gain, that causes the signal to drop to zero as it fully replaces the ${\rm Ar-CO}_2$.
The nitrogen is then displaced with the operating gas, allowing for the time dependence of the gain onset to be measured.
As the ${\rm Ar-CO}_2$ begins to replace the ${\rm N}_2$, the gain is gradually restored, and the source will once again induce a signal in the straws. 
When all the nitrogen has been purged from the panel, the doublet signals will reach their maximum level.

The best performance of the replacement procedure is achieved when gases begin to be replaced in the straws simultaneously, enabling uniform and comparable measurements across the panel.
This is accomplished by opening the auxiliary valve on the inlet side cover and flushing with the replacement gas.
Since the impedance of the auxiliary valve is lower than that of the straws, most of the introduced gas fills the inlet side rather than the straws.
The intended flow path when the auxiliary valve on the inlet side is open is illustrated by the solid arrows in Figure~\ref{fig:Fe55-openvalve}.
After allowing the replacement gas to flow for a suitable time to fill the inlet side ($\sim$\(500\,\text{s}\)), the auxiliary valve is closed.
This causes the introduced gas to begin displacing the previous gas in all straws.



\begin{figure*}[!t]
  \centering
  \begin{subfigure}{0.45\textwidth}
    \centering
    \includegraphics[width=0.9\linewidth]{ 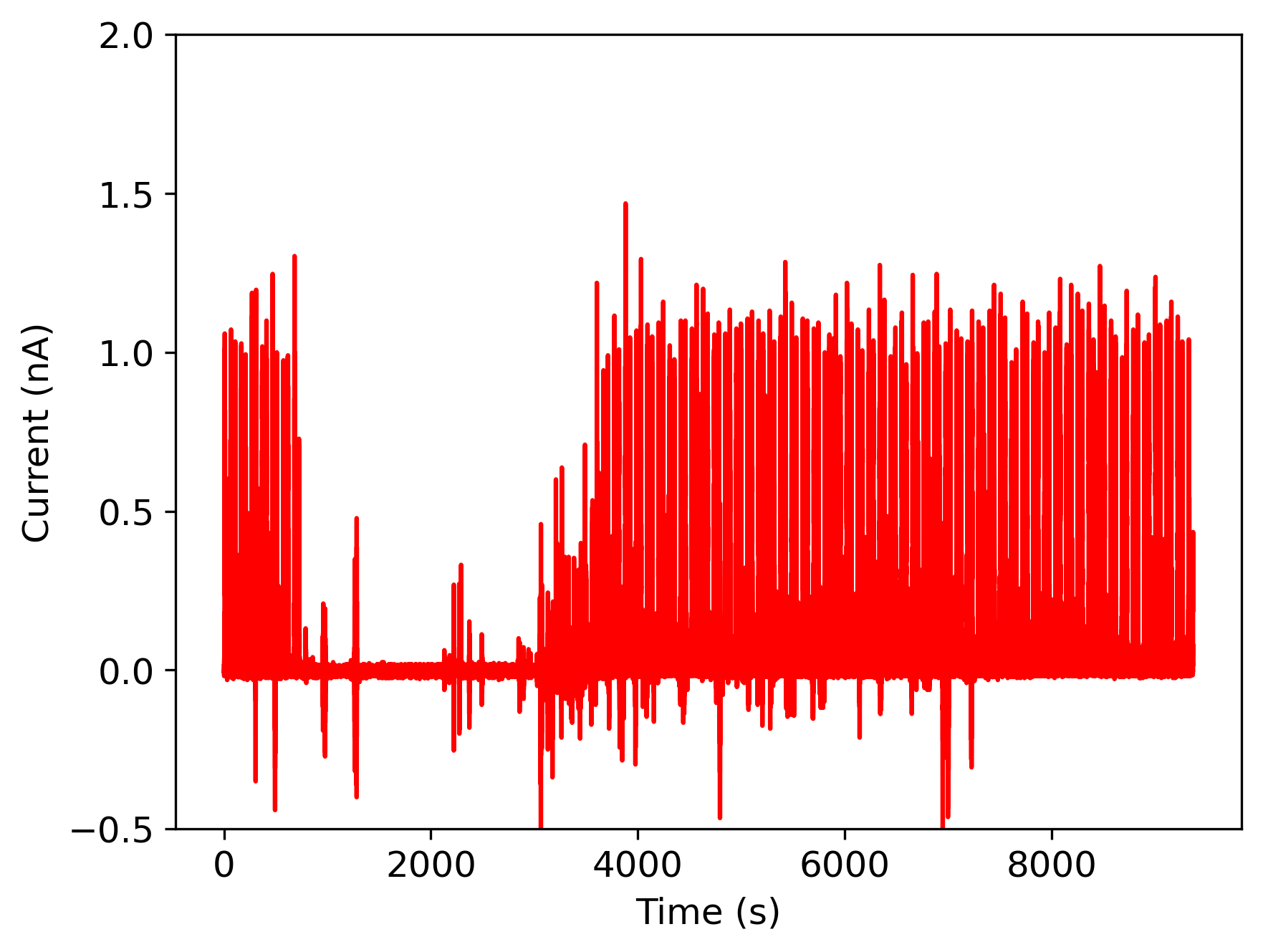}
    \caption{\label{fig:rawdata}}
  \end{subfigure}
  \begin{subfigure}{0.45\textwidth}
    \centering
    \includegraphics[width=0.9\linewidth]{ 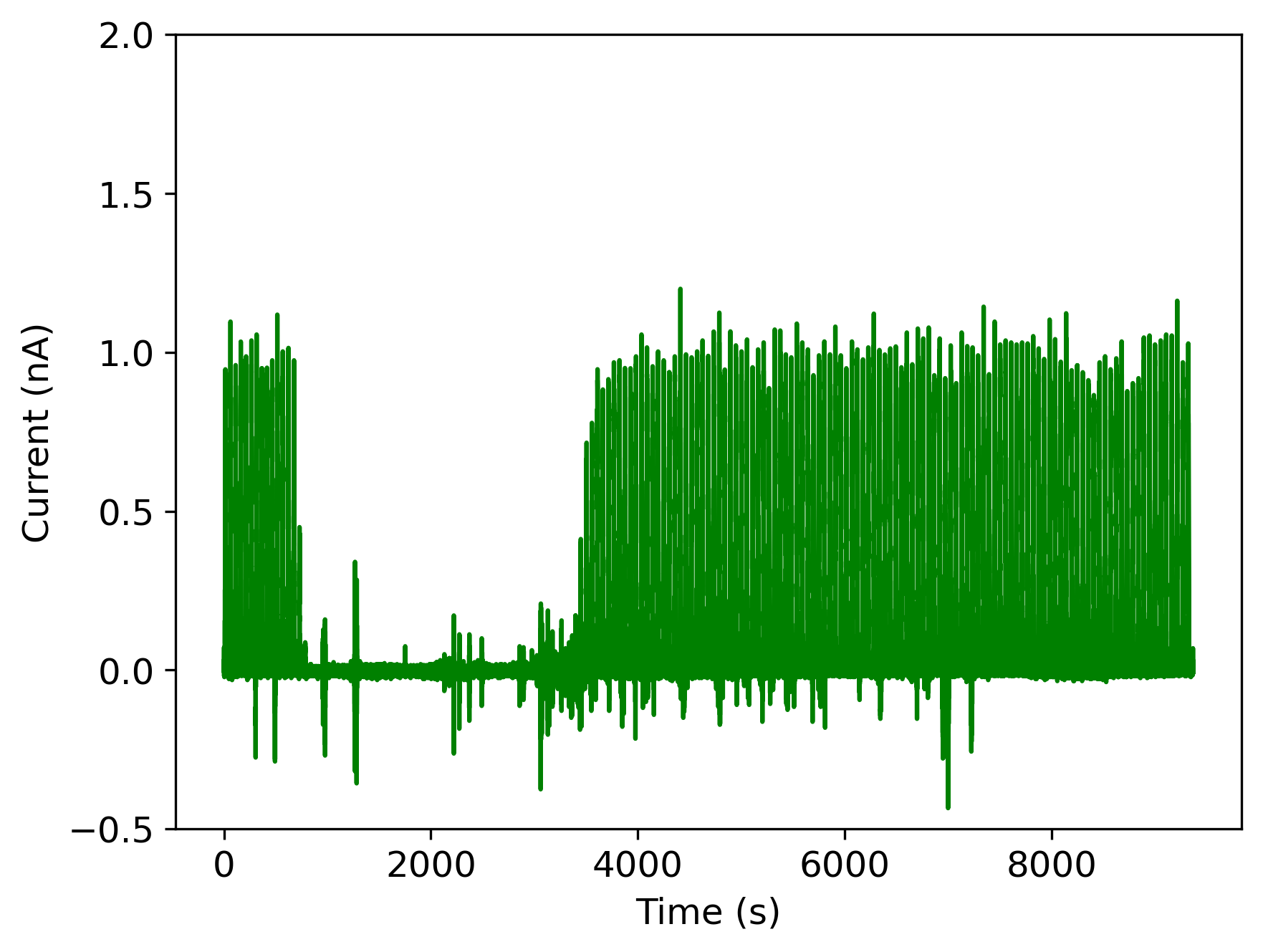}
    \caption{\label{fig:smoothdata}}
  \end{subfigure}
  \caption{\label{fig:compare} Example current data from a straw doublet during a rise time study. (a) Raw current data. (b) Smoothed current data.}
\end{figure*}

\begin{figure}[!b]
  \centering
  \includegraphics[width=0.45\textwidth]{ 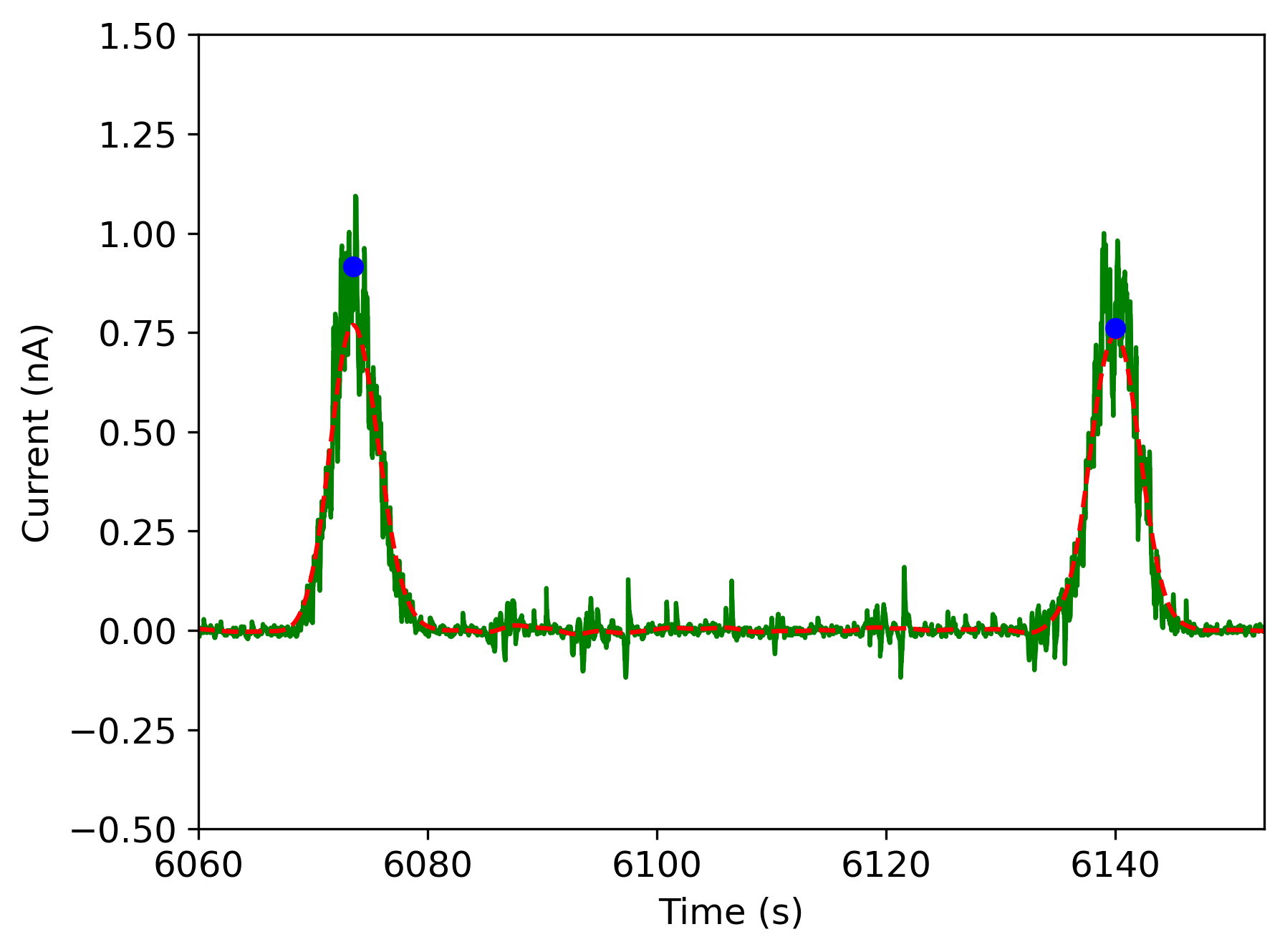}
  \caption{\label{fig:gausfit} Measurements of Gaussian peaks for two source passes in a doublet. The solid line represents smoothed data, the dashed line displays processed data, and the point marks the current peak value.}
\end{figure}

\begin{figure}[!t]
  \centering
  \includegraphics[width=0.45\textwidth]{ 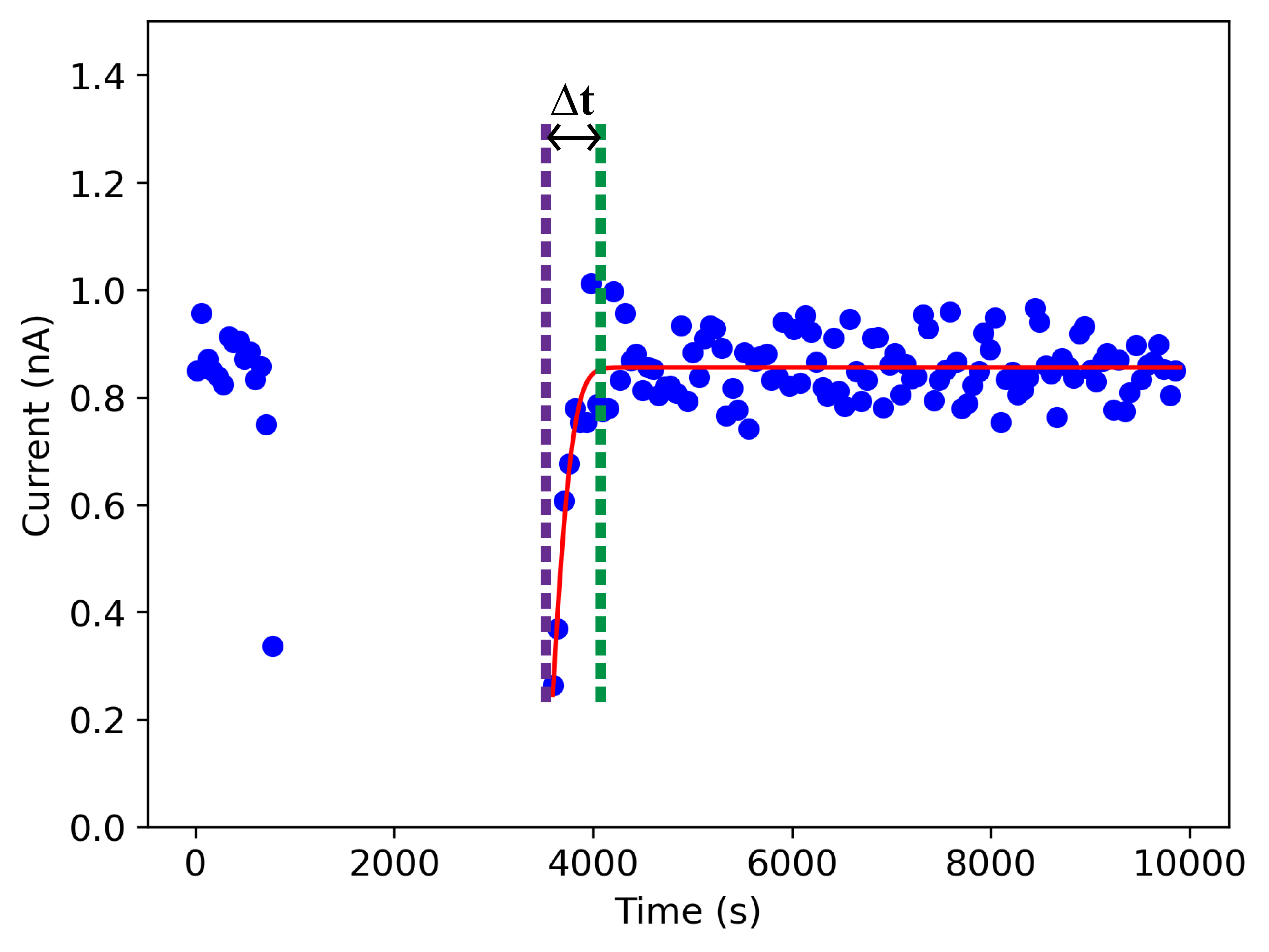}
  \caption{\label{fig:erffit} Processed data for one test run of a single doublet. The points correspond to each peak, as measured in the Gaussian fit. The solid line fit to the peak data is an error function fit for the region of rising current. The first dashed line indicates when the auxiliary valve is closed at $t_0$, and the second dashed line marks when the measured current reaches 90$\%$ of the fit maximum at $t_1$. The difference between these represents the rise time, $\Delta t$.}
\end{figure}

\subsection{Data Analysis Method}

Measuring the current over time from each doublet during the test allows us to monitor the response to the gas exchange.
Figure~\ref{fig:compare} presents representative test data, with current measurements taken every 0.1~s.
In the regions of high current, the regular peaks occurring every 60~s are due to the \(\mbox{}^{55}\mathrm{Fe}\) source passing by the doublet. 
These peaks are approximately Gaussian, and the area under the curve represents the total charge collected. 
At the start time, the panel has been filled with ${\rm Ar-CO}_2$ and the source is sweeping.
After roughly \(1000\,\text{s}\), the gas mixture has been entirely replaced by the ${\rm N}_2$, causing the straw current to drop to zero.
At approximately \(3000\,\text{s}\), the gain is restored as the ${\rm Ar-CO}_2$ refills all the straws.

The raw current data shown in Figure~\ref{fig:rawdata} is inherently noisy, consisting of various other ionization phenomena.
To suppress noise, a running average algorithm is applied to produce the smoothed signal shown in Figure~\ref{fig:smoothdata}.
If a data point exceeds its neighbors by more than a threshold of \(0.025\,\text{nA}\), which is close to the average current, it is replaced by the average of the adjacent points. 
Since the window size is small compared to the speed of the source, the \(\mbox{}^{55}\mathrm{Fe}\) peaks remain unaffected while the nearly instantaneous fluctuations are filtered out.

\begin{figure}[!t]
  \centering
  \begin{subfigure}{0.48\textwidth}
    \centering
    \includegraphics[width=0.9\linewidth]{ 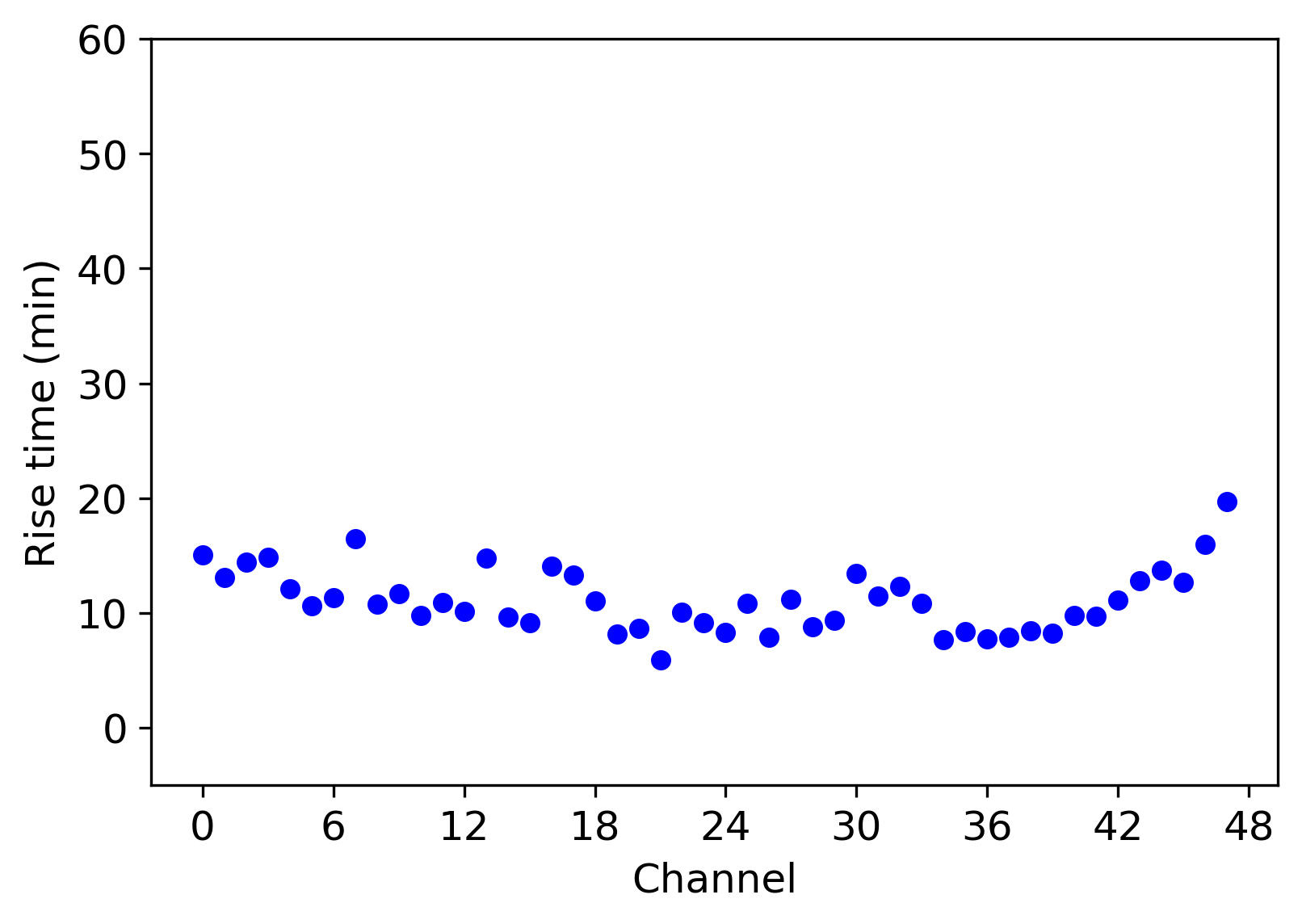}
    \caption{\label{fig:deltatime_plot}}
  \end{subfigure}
  \begin{subfigure}{0.48\textwidth}
    \centering
    \includegraphics[width=0.9\linewidth]{ 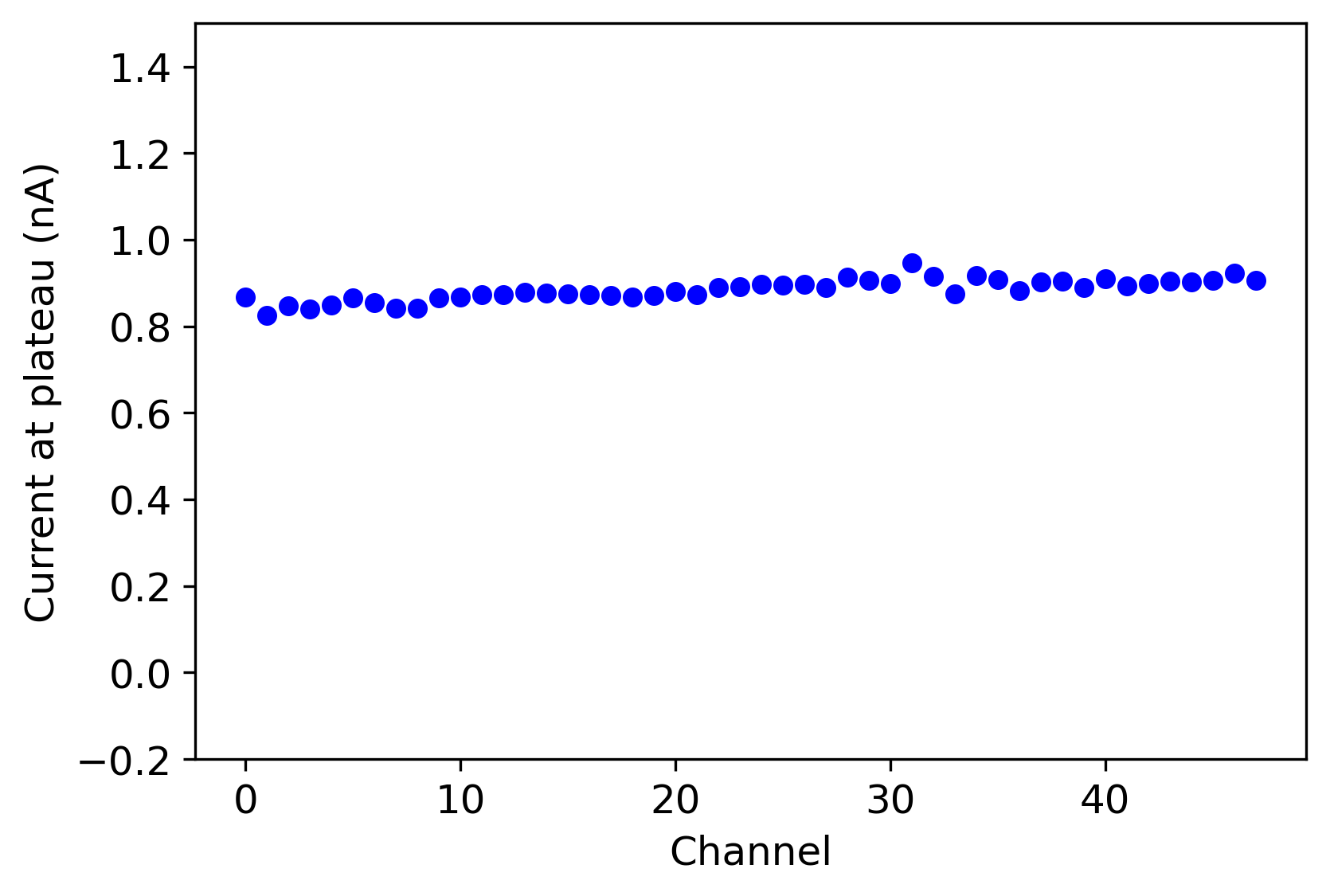}
    \caption{\label{fig:gain_plot}}
  \end{subfigure}
  \caption{\label{fig:result} Typical results from one panel for the straw tube quality control test. (a) Rise times for each straw doublet. (b) Current gain measurements, the maximum value of the error function fit, for each straw doublet.}
\end{figure}

\begin{figure}[!tb]
  \centering
  \includegraphics[width=0.45\textwidth]{ 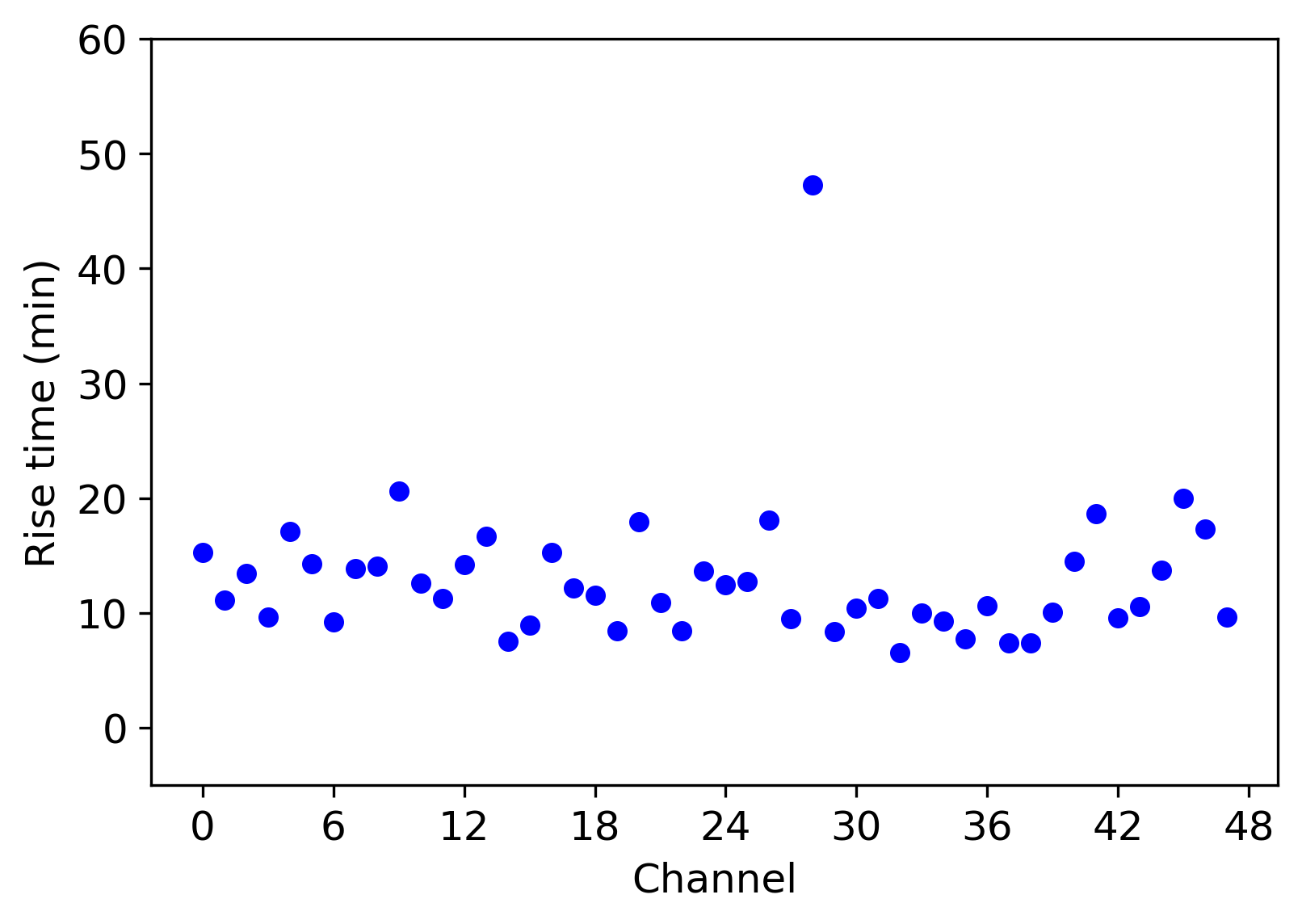}
  \caption{\label{fig:blockedstraw_panel} Example of a panel with a potentially blocked doublet connected to channel 28.}
\end{figure}

\begin{figure*}[tb]
  \centering
  \begin{subfigure}{0.405\textwidth}
    \centering
    \includegraphics[width=\linewidth]{ 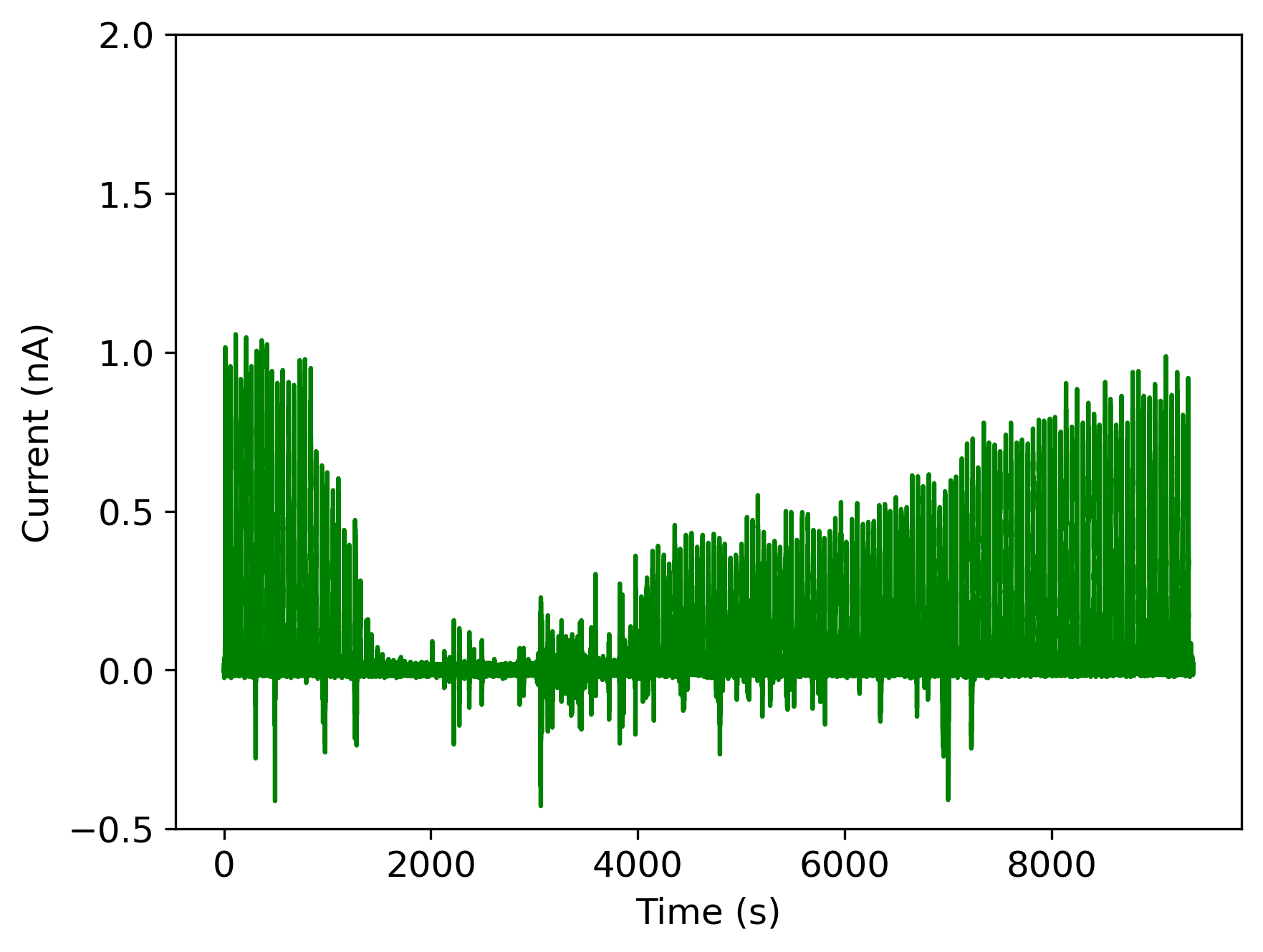}
    \caption{\label{fig:blockedstraw_channel}}
  \end{subfigure}
  \begin{subfigure}{0.405\textwidth}
    \centering
    \includegraphics[width=\linewidth]{ 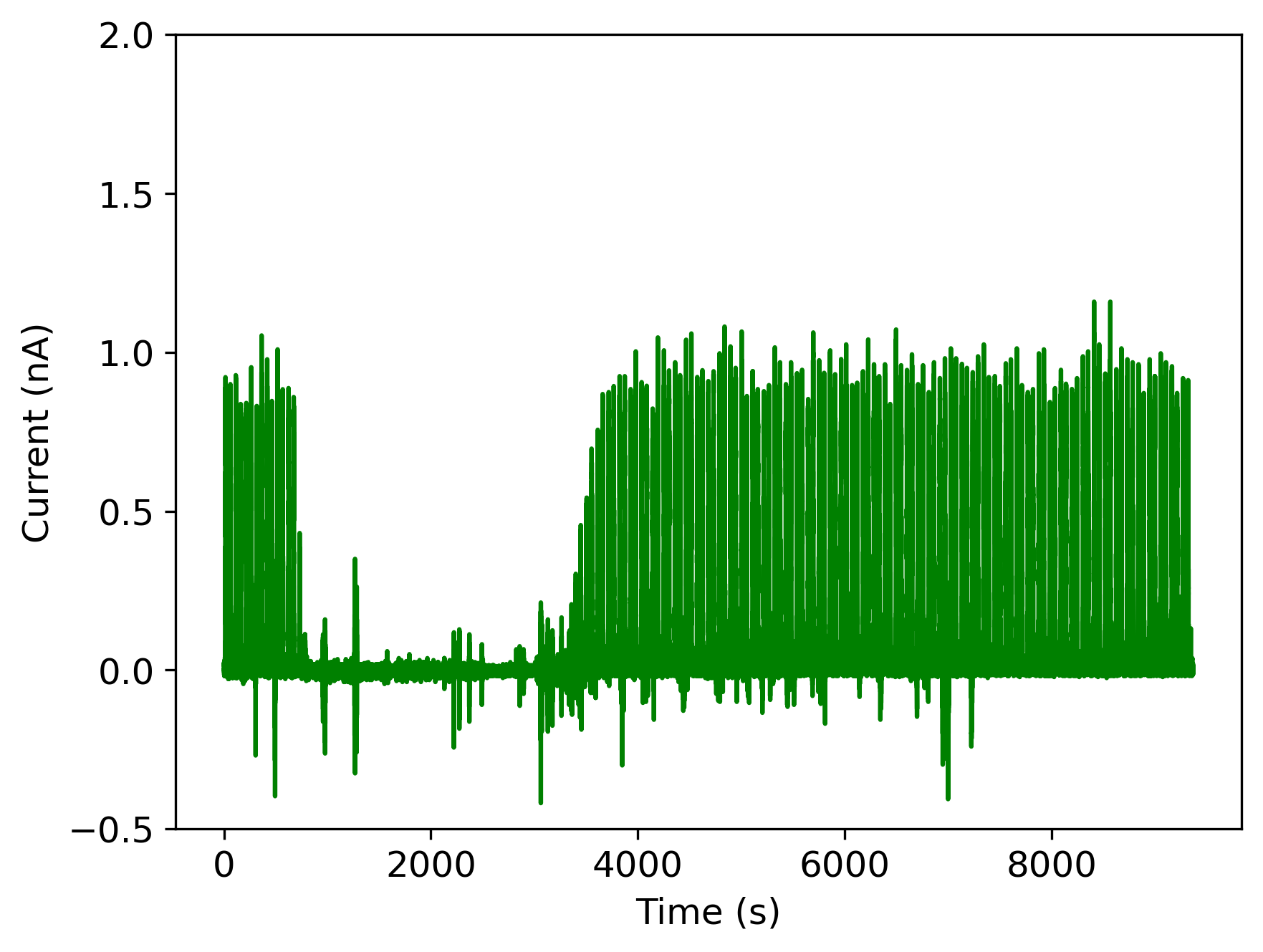}
    \caption{\label{fig:blockedstraw_channel_repair}}
  \end{subfigure}
  \caption{\label{fig:result1} Example of a partially blocked doublet connected to channel 28. (a) Smoothed data before repair. (b) Smoothed data after repair.}
\end{figure*}

To remove any remaining fluctuations, the smoothed data is processed using the open source package \texttt{SciPy}'s Gaussian filter \allowbreak (\texttt{gaussian\_filter}).
With this processed data, we identify each peak induced by the radiation source as it passes each doublet (Fig.~\ref{fig:gausfit}).
We utilize \texttt{SciPy}'s \allowbreak \texttt{find\_peaks} function to locate the center of the Gaussian peaks in time.
The recorded peak value is calculated from the smoothed data over a 0.5~s time window around each identified peak position in the processed data.
Only fits with a peak large enough for a reliable measurement (\textgreater\(0.05\,\text{nA}\)) are recorded.

For each doublet, the extracted peak height as a function of time is fit using an error function, which models the gain onset.
Figure~\ref{fig:erffit} shows an example error function fit, which is used to characterize the rise time and the gain.
The gain onset time or ``current rise time'' for a doublet is the time interval between auxiliary valve closure and when the gain is restored; that is, when the current reaches $90\%$ of the error function maximum.
The gain is defined as the maximum current for the error function.

The error function fit results are evaluated across all 48 cathode channels, each corresponding to a doublet in the panel.
Data from a representative panel is shown in Figure~\ref{fig:result}.
Figure~\ref{fig:deltatime_plot} shows the rise time and Figure~\ref{fig:gain_plot} shows the current gain.

In principle, the rise time should increase with straw length as the gas volume increases.
However, in our panel design the gas inlet is located nearest to the longest straws (Fig.~\ref{fig:Fe55-openvalve}). 
This results in a gas pressure differential that slightly increases the rise time values for the shorter straws, which are farthest from the inlet.
This relationship can be observed in Figure~\ref{fig:deltatime_plot}, where channel~0 corresponds to the longest doublet and channel~47 corresponds to the shortest.

\subsection{Evaluating Results}

\begin{figure}[!b]
  \centering
  \includegraphics[width=0.405\textwidth]{ 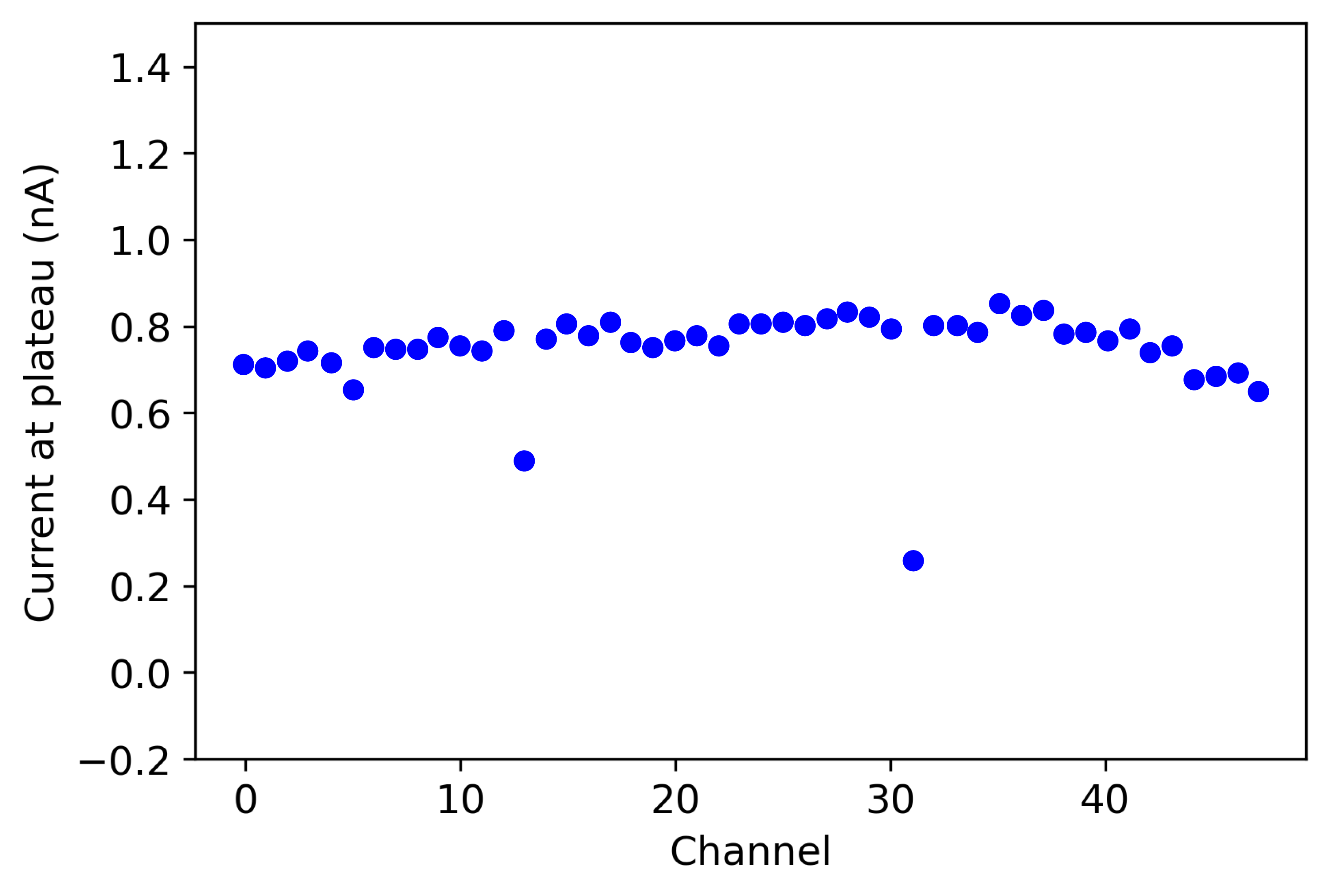}
  \caption{\label{fig:gainblocked} Example gain plot for a panel with two potentially blocked doublets connected to channels 13 and 31.}
\end{figure}

To diagnose any issues, we evaluate the rise time and gain measurements for each panel.
A straw with a restriction takes longer to exchange gas, resulting in a longer rise time.
Thus, any doublet with an unusually long rise time is flagged as potentially having a flow restriction.
In the example panel shown in Figure~\ref{fig:blockedstraw_panel}, channel~28 exhibits a rise time that is more than double that of any other doublet.
Therefore, this channel is likely to have a flow restriction.

Doublets with potential flow restrictions are re-examined using their smoothed current data.
At this level, the effects of a prolonged gain onset can be confirmed.
Figure~\ref{fig:blockedstraw_channel} shows the smoothed data for a partially blocked doublet, a stark difference from the typical signal in Figure~\ref{fig:smoothdata}.
When refilled, the gain onset time is significantly longer than usual, indicating the presence of residual ${\rm N}_2$.
This behavior confirms there is an obstruction.


The rise time and gain measurements only reveal which doublets have flow restrictions.
To identify which straw within a doublet is blocked, a soft and extremely thin probe is inserted to locate the physical obstruction.
Then, an attempt is made to repair the affected straw by drilling out the gas passages at the terminations to clear stray hardened epoxy and remove any debris.

The panel is retested and the new smoothed results are shown in Figure~\ref{fig:blockedstraw_channel_repair}.
The data shows a regular pattern of signal reduction during gas exchange, between $\sim$\(1000\,\text{s}\) and $\sim$\(3000\,\text{s}\).
The rise time is now consistent with the results shown in Figure~\ref{fig:deltatime_plot}, indicating that the repair was successful.

Gain measurements provide a complementary evaluation of performance alongside rise time and can be used to identify blockages.
A suppressed gain indicates a lower current plateau and, therefore, an obstructed flow in the doublet.
In the panel shown in Figure~\ref{fig:gainblocked}, channels 13 and 31 display a lower gain compared to other doublets.
Such channels are then re-examined using the same approach as the rise time analysis to confirm a blockage.

\section{Conclusion}

The straw tube tracker provides precise momentum measurements that are central to the Mu2e experiment at Fermilab.
To ensure its performance, a quality control test was developed that assesses gas flow through all straws in each panel.

Over the course of two years, this test was performed on all 11,280~doublets (covering all tracker panels and spares) and flow restrictions were identified in 219~($1.94~\%$).
After repairs, 164~doublets~($74.9~\%$) restored their performance.
At the single straw level, $0.95~\%$ of straws were found to have a blockage.
Of these, $76.3~\%$ blocked straws were recovered, with the remainder failing due to wire breakage during repairs.
As a result, all operational tracker panels meet the gas flow requirements, with only a small amount of straws remaining problematic.

This technique has improved the uniformity and coverage of the tracker, ensuring proper detection resolution.
Gain onset measurements of this form can be adapted for use in other straw tube detectors to improve their performance.


\section{Acknowledgments}

We are grateful for the vital contributions of the Fermilab staff and the technical staff of the participating institutions. This work was supported by the US Department of Energy; the Istituto Nazionale di Fisica Nucleare, Italy; the Science and Technology Facilities Council, UK; the Ministry of Education and Science, Russian Federation; the National Science Foundation, USA; the National Science Foundation, China; the Helmholtz Association, Germany; the Julian Schwinger Foundation, USA; and the EU Horizon 2020 Research and Innovation Program under the Marie Sklodowska-Curie Grant Agreement Nos. 858199, 101003460, and 101006726. This document was prepared by members of the Mu2e Collaboration using the resources of the Fermi National Accelerator Laboratory (Fermilab), a U.S. Department of Energy, Office of Science, Office of Science, Office of High Energy Physics HEP User Facility. Fermilab is managed by FermiForward Discovery Group, LLC, acting under Contract No. 89243024CSC000002.
We acknowledge that authors at CUNY were supported under U.S. Department of Energy contract DE-SC0019027.






\end{document}